\documentstyle[multicol,aps]{revtex}
\def\baselinestretch{1.1}

\renewcommand{\narrowtext}{\begin{multicols}{2} \global\columnwidth20.5pc}
\renewcommand{\widetext}{\end{multicols} \global\columnwidth42.5pc}

\begin{document}

\title{\Large\bf  On area and entropy of a black hole }
\author{ \large\bf A. Alekseev$^{\dag}$, 
A. P. Polychronakos$^{\dag *}$, M. Smedb\"ack$^{\dag}$ \\[2mm] }

\address{\noindent $^{\dag}$  
Institute for Theoretical Physics, Uppsala University,
 Box 803, S-75108, Uppsala, Sweden\\
$^*$ Theoretical Physics Dept.,  University of Ioannina,
45110 Ioannina, Greece\\
e-mails: alekseev@teorfys.uu.se, poly@teorfys.uu.se,
mikael@teorfys.uu.se \\
 }

\date{April 2000} 

\maketitle

\begin{abstract}
\noindent

We consider a model of a black hole consisting of a number
of elementary components. Examples of such models occur in
the Ashtekar's approach to canonical Quantum Gravity
and in M-theory. We show that treating the elementary components
as completely distinguishable leads to the area law for the 
black hole entropy. Contrary to previous results,
we show that no Bose condensation occurs,
the area has big local fluctuations  and that
in the framework of canonical Quantum Gravity the
area of the black hole horizon is equidistantly quantized.

\vspace{1mm}
\noindent
PACS-98: 04.70.Dy, 11.15.-q
\end{abstract}

\narrowtext

%\section{Introduction}

In the framework of Quantum Gravity black holes are treated as
quantum objects. As such, they are characterized by quantum numbers:
mass, electric charge, angular momentum {\em etc}.
For Schwarzschild black holes (neutral, nonrotating)
the only quantum number which is left is the mass $M$. It is related to
the area $A$ of the black hole horizon by formula,
$A= 16 \pi G^2 M^2/c^4$, where $G$ is the gravitational constant. Important questions in
black hole physics are what the spectrum of $A$ looks like and what the 
degeneracies of states are for a given value of $A$. 

In the absence of a definitive Quantum Gravity theory, the answers
to these questions depend on the model of a black hole. In several
approaches one assumes that a black hole consists of elementary
components contributing additively to its area. For instance, in the 
Ashtekar's approach (see {\em e.g.} \cite{Gambinibook,Ashtekarbook})
these are Wilson lines of the Ashtekar's connection $A_\mu^a$. In the $M$-theory
approach  \cite{D0} these are D$_0$-branes. Each of the elementary
components can be in a number of states.
An elementary component in the state $n$ gives a contribution $A_n$ to the
total area, and has a degeneracy $g(n)$.
The eigenvalues of the area operator then acquire the form
$A= \sum_{n=1}^\infty  N_n A_n $,
where $N_n$ is the number of elementary components in the state $n$.
The multiplicity of states with area $A$ is $\Omega (A)$ and its asymptotics 
when $A$ is macroscopically large defines the entropy of the black hole,
$S(A)  = k \ln \Omega(A)$.

The spectrum of the area $A$ and the behaviour of the entropy
$S(A)$ depend on the spectrum of the elementary area $A_n$, the elementary 
multiplicities $g(n)$ and the degree of distinguishability of the components. 
We will discuss
the form of $A_n$ in the framework of canonical Quantum Gravity 
\cite{Gambinibook,Ashtekarbook}.
The classical result in this field is that the index $n$ can be 
associated with a spin $j$ of an $SU(2)$ representation. Then, $j$ takes
integer and half-integer values and according to \cite{RS_spectrum} 
$A_j= \sqrt{j(j+1)}$ in some units. We argue that instead 
\begin{equation}  \label{Aj}
A_j=j+ \frac{1}{2}.
\end{equation}
In particular, this implies that the spectrum of the area operator is
equidistant. Such a situation was first considered by Bekenstein and
Mukhanov \cite{spectroscopy}. Their analysis shows that such a spectrum 
implies certain specific properties of the black hole radiation: 
there exists an energy quantum,
$\hbar \omega_0$, and the energy can be radiated only in integer multiples
of this energy quantum.

Furthermore, we will consider the issue of black hole entropy. It will be 
seen that the {\it area law} $S \propto A$ is to a large extent generic.
A more subtle issue is what the statistically preferrable
state of the system is. The analysis of \cite{entropy} in the case of Ashtekar
gravity and of \cite{CM} in the case of black holes composed
of D$_0$-branes
showed that the most probable configuration
has all the elementary components in the same multiplet. This can be
viewed as Bose condensation. 
We show instead that for completely distinguishable elementary
components one obtains a Gibbs distribution for
the $N_n$'s with
many of them nonvanishing in the most probable configuration.
Our result implies that the area operator has strong fluctuations,
$\Delta A \sim \sqrt{A}$, whereas in the case of Bose condensation
the fluctuation of area would have been strongly suppressed.

Next we turn to the issue of the area spectrum in canonical
Quantum Gravity.
One of the virtues of Ashtekar's formulation 
is that instead of the metric $g_{\mu \nu}$ one deals with
the gauge field $A_\mu^a$ \cite{Gambinibook,Ashtekarbook}. 
In the Euclidean gravity the gauge
group is $SU(2)$ and the isotopic index $a$ takes values
$1,2,3$ corresponding to three Pauli matrices. In the Hamiltonian 
formulation one uses the spatial components $A_i^a$ as generalized
coordinates and the components of the electric field,
$E_i = \partial_0 A_i - \partial_i A_0 - [A_0, A_i]$ as conjugate momenta.
It is convenient to view the gauge field  as a 1-form $A=A_i dx^i$, and the 
conjugate momenta $E_i$ as components of a 2-form, $E=\frac{1}{2}
\varepsilon_{ijk} E_i dx^j dx^k$.

The physical Hilbert space is obtained by imposing the Gauss law
constraint, the diffeomorphism constraint and the Hamiltonian
constraint (the Wheeler-DeWitt equation). While constructing
wave functionals which annihilate all those constraints proves to
be a difficult task \cite{Gambinibook}, one believes that the spectrum of the 
area operator is already captured by studying the `kinematical
Hilbert space' containing the Wilson lines \cite{Baez},
\begin{equation} \label{Wilson}
W^j_\Gamma(A) = {\rm Tr} P \exp(\int_\Gamma A^a T^a_j ),
\end{equation}
where $\Gamma$ is a closed contour, $j$ is a positive integer or half-integer and
$T_j^a$ are generators of the Lie algebra $su(2)$ in the representation with
spin $j$.

In terms of the Ashtekar's variables, the area operator 
corresponding to the 2-dimensional surface $\Sigma$ acquires
the form \cite{RS_spectrum},
\begin{equation}  \label{area}
{\mathcal A}_\Sigma = \int_\Sigma d^2x \sqrt{{\rm Tr} \, E^2},
\end{equation}
where the integrand is the natural density which can be integrated
over the 2-dimensional surface $\Sigma$.
The Wilson lines $W_\Gamma^j$ are eigenstates of the operators
${\mathcal A}_\Sigma$, at least when all the intersections of $\Gamma$ and
$\Sigma$ are transversal. Moreover,  each intersection gives a 
contribution $A_j$ which depends only on the spin $j$ of the Wilson
line. It is our next task to compute the numbers $A_j$.

Canonical quantization suggests that in the $A$-representation
the conjugate momentum $E_i^a$ act as derivatives,
$E_a^i(x)  = - i  \delta/\delta A_a^i(x)$.
Ignoring the singularity arising from the coincident arguments 
in the expression $\sqrt{E_i^a(x)E^a_i(x)}$, one easily obtains
$$
{\mathcal A}_\Sigma W_\Gamma(A) = \sum_p {\rm Tr} \sqrt{T^a_j T^a_j} 
P \exp(\int_p A_i^a T_j^a dx_i),
$$
where $p$ are the intersection points  of $\Gamma$ and $\Sigma$,
and $\int_p$ stands for the integration over $\Gamma$ with the starting
point at $p$. The expression $T^a_j T^a_j$ is proportional to the 
unit matrix with coefficient the quadratic Casimir $c_j=j(j+1)$ of $SU(2)$.
This observation led \cite{RS_spectrum} to the conclusion that $A_j=\sqrt{j(j+1)}$.
We shall show instead that formula (\ref{Aj}) holds true.

The area operator  $(E_i^a(x)E^a_i(x))$ contains a product of 
two fields $E_a^i(x)$ at the same point and potentially needs a 
regularization. The answer $c_j=j(j+1)$ is certainly correct to the 
leading order in $j$ which might be used as  the parameter
in the semi-classical expansion. However, there might be quantum 
corrections to this formula similar to the shift by $\hbar \omega/2$
in the energy spectrum of the harmonic oscillator 
$E_n=\hbar \omega(n+\frac{1}{2})$.

First, let us mention that the quantization 
problem for polynomials
of commuting variables $[t^a,t^b]=0$ which after quantization
$t^a \mapsto T^a$ acquire Lie algebra commutation relations
$[T^a,T^b]=f_{abc}T_c$ has a long history in mathematics. 
In fact, there is a universal {\em quantization map} ${\mathcal Q}$ which 
was discovered by Harish-Chandra for semi-simple
Lie algebras and then generalized by Duflo \cite{Duf} to the case of arbitrary
Lie algebras. The main property of ${\mathcal Q}$ is that given two
Casimir elements $\alpha$ and $\beta$ (a particular example of
a Casimir element is the quadratic Casimir $T^aT^a$) the product
of quantizations ${\mathcal Q}(\alpha) {\mathcal Q}(\beta)$ coincides with
the quantization of the product,  ${\mathcal Q}(\alpha \beta)$.
In more detail,
$$
{\mathcal Q} (\alpha) = {\rm Sym} \, \left[ {\rm det}\left( \frac{\sin(x)}{x} 
\right)_{x= T^a_{{\rm ad}} \frac{\partial}{\partial t^a} } \alpha(t) \right],
$$
where $T^a$ are generators in the adjoint representation, and
${\rm Sym}$ stands for the symmetric quantization map,
{\em e.g.} ${\rm Sym} (t^at^b) = \frac{1}{2}(T^aT^b +T^bT^a)$.
For instance, in the case of the Lie algebra $su(2)$,
${\mathcal Q}= {\rm Sym}\, (1 + \frac{1}{12}\frac{\partial^2}{\partial t^a \partial t^a} + \dots)$,
where $\dots$ stands for the terms containing higher derivatives in $t^a$.
When applying ${\mathcal Q}$ to $c=t^at^a$ one obtains 
${\mathcal Q}(c)=T^aT^a + \frac{1}{4}$ with the eigenvalue
${\mathcal Q}(c)_j = (j+\frac{1}{2})^2$ in the spin $j$ 
representation. Note that the minimal value of the Casimir
corresponding to the (infinite-dimensional) representation
of spin $j=-\frac12$ is equal to zero.

Second, instead of using the $A$-representation one can rewrite
the wave functional $W_\Gamma(A)$ in the $E$-representation
by means of the functional Fourier transform,
\begin{equation} \label{Fourier}
\tilde{W}_\Gamma(E)=\int {\mathcal D}A \, e^{i \int {\rm Tr}\,  (EA)} \, W_\Gamma(A).
\end{equation}
This expression can be simplified using the geometric quantization formula
for the Wilson line \cite{NR,AFS},
\begin{equation} \label{geom}
W_\Gamma(A) = \int {\mathcal D}g \, e^{i  \int_\Gamma {\rm Tr}
 (\tau g^{-1}\partial_s g ds + A  (g\tau g^{-1}) )},
\end{equation}
where the auxiliary field $g$ is a group valued function on the
contour $\Gamma$, $\tau$ is a constant diagonal matrix, and $s$ is a 
parameter along $\Gamma$. The right hand side of (\ref{geom})
is well-defined only when the eigenvalues of $p$ are integers or
half-integers \cite{AFS}.
Putting together the functional 
Fourier transform (\ref{Fourier}) and the geometric quantization (\ref{geom})
yields,
\begin{equation} \label{WE}
\tilde{W}_\Gamma(E) = \int dg \delta(E - g\tau g^{-1}  \delta_\Gamma),
\end{equation}
where $\delta_\Gamma$ stands for the $\delta$-function supported on
$\Gamma$. Equation (\ref{WE}) suggests that $E$ vanishes outside
$\Gamma$. On the contour, we obtain $\sqrt{{\rm Tr} E^2} = 
\sqrt{{\rm Tr} \tau^2} \delta_\Gamma$. Hence, the contribution to the
area operator coming from the transversal intersection of $\Gamma$
and $\Sigma$ is given by $A_j=\sqrt{{\rm Tr} \tau^2}$. The relation between $\tau$
and $j$ depends on the regularization of the functional integral over
$g$. Using the regularization of \cite{NR} one obtains 
$A_j=j+\frac{1}{2}$ (the regularization of \cite{AFS} yields $A_j=j$).

Third, the same problem of regularization of the ${\rm Tr} E^2$ operator
arises in  2-dimensional Yang-Mills theory as considered by Witten
\cite{Witten}.
There, the value of the regularized quadratic Casimir has an influence
on the results for the volumes of the moduli spaces of flat connections
on Riemann surfaces. The same volumes can be computed by 
a different mathematically rigorous procedure (see {\em e.g.} \cite{MW}).
Once again, the correct regularization of the Casimir element
is $c_j=(j+\frac{1}{2})^2$ \cite{Witten}.

Finally, upon proper use of the gauge constraints, the problem 
reduces to geodesic
motion of a particle on the group manifold, in our case $SU(2)$. 
The operator
Tr $E^2$ becomes the Hamiltonian of the particle. Quantum 
mechanically, 
for motion on a curved manifold, it is known that in addition to the
Laplacian 
(which would reproduce the standard Casimir)
a term proportional to the scalar curvature of the manifold should
be added,
with the coefficient fixed to $\frac{1}{8}$ in order to reproduce the right
conformal
properties. In our case, this term is a constant and contributes 
the extra $\frac{1}{4}$.
Alternatively, one can reduce the system even further to the dynamics
of the 
eigenvalues of the group element. It is known that these
behave as 
free fermions on a circle, and the extra contribution is seen to be 
the ground
state Fermi energy of the system.

We conclude that all the evidence is in favour of the formula (\ref{Aj})
for the spectrum $A_j$ of the elementary objects. This implies the
Bekenstein-Mukhanov type discrete spectrum of the black hole area,
and the discrete spectrum of the black hole radiation.

%\section{Black Hole Entropy}\label{BHE}

We now turn to the black hole thermodynamics.
We consider a quantum black hole as
consisting of a large number of identical 
elementary components. Examples of such elementary components
are the Wilson lines in the Ashtekar's gravity and
D$_0$-branes in M-theory.
Different viewpoints on the degree of 
distinguishability of these components will be discussed, and the entropy
will be calculated in each case. 
Viewing the set of elementary components
as the constituents of a grand canonical system, 
thermodynamics determines the entropy $S$ of 
the black hole, 
to be given by
\begin{equation} \label{entropy}
  S=k(A \beta + \ln Z),
\end{equation}
where $k$ is the Boltzmann constant, $A$ is the conserved black hole area, 
$\beta$ is a temperature-like parameter dual to the area and $Z$ 
is the partition function,
\begin{equation} \label{partitionfcn}
  Z = \sum_{A} \Omega(A) e^{-\beta A} = \sum_{A,N} \Omega(A,N) e^{-\beta A}.
\end{equation}
Here $\Omega(A)$ is the multiplicity of states of area A, and 
$\Omega(A,N)$ is the multiplicity of state of area A and $N$ elementary
components. The 
summation is over all possible areas and number of components. 

To proceed, we will consider the elementary components as identical but {\it completely 
distinguishable} independent quantum systems. Then the partition 
function becomes
\begin{equation} \label{partitionfcn2}
  Z = \sum_{N} Z_{1}^{N}=\frac{1}{1-Z_1},
\end{equation}
where $ Z_1 = \sum_{A} \Omega(A,1) e^{-\beta A}$ is the partition function of 
a single component.
The above expression for $Z$ implies that $\beta$ can never be less than
the {\it Hagedorn temperature} parameter $\beta_o$, fixed by the relation
$Z_1 ( \beta_o ) = 1$.

From equation (\ref{partitionfcn}) the area is related to the 
partition function by formula,
$$
A=-\frac{d \ln Z}{d\beta} = -\frac{1}{1-Z_1} \cdot \frac{dZ_1}{d\beta}.
$$
Restricting our considerations to macroscopic areas $A \gg 1$ only, this relation 
implies that $1-Z_1 \sim \frac{1}{A}$, and thus $\beta \to \beta_o$ and $\ln Z$ 
grows only as $\ln A$. The dominant
contribution to the entropy will therefore be given by $S=k\beta_o A$. 
The entropy is always proportional to the area. This is a completely general 
result, valid for any system consisting of distinguishable components.

To fix the proportionality constant, we restrict our considerations to the case 
of the area spectrum given by equation (\ref{Aj}),{\em i.e.} $A_n=n$, 
$n=1,2,3,...$, and the degeneracy function $\Omega(A,1) = 
\Omega(n,1)=g(n)=n$. Note that the states enumerated by the spin quantum 
number $j$, taking integer or half-integer values, are now enumerated by $n=2j+1$, 
a positive integer. We obtain
$Z_1 = (2 \sinh\frac{\beta}{2})^{-2}$.
The partition function diverges at the Hagedorn temperature 
$\beta=\beta_0=\ln\frac{3+\sqrt{5}}{2}$ and the area becomes macroscopic as 
$\beta \to \beta_0$. In that limit the entropy becomes
\begin{equation} \label{final_entropy}
  S=\frac{kA}{4\pi\gamma l_p^2} \ln \left( \frac{3+\sqrt{5}}{2} \right),
\end{equation}
where we have acknowledged that the area is measured in units of 
$4\pi\gamma l_p^2$, where $l_p$ is the Planck length and $\gamma$
is the Immirzi parameter \cite{Immirzi}.

Equation (\ref{final_entropy}) represents our final result for the entropy of a 
black hole. The validity of this derivation depends crucially on two important 
claims: that equation (\ref{Aj}) gives the true area spectrum and that the area 
constituents, i.e. the edges in the Ashtekar's gravity approach, are completely 
distinguishable. Attempts to justify the former claim have already been made. Let us now turn to the latter.

The elementary components of a black hole could {\it a priori} be either 
{\it identical}, {\it partially distinguishable} or {\it completely distinguishable}.
Completely distinguishable components would mean that: 
\noindent
(1) There is a difference between assigning two given spin values $j_1$ and
$j_2$ to two different edges according to (edge 1, edge 2)$=(j_1,j_2)$ and 
(edge 1, edge 2)$=(j_2,j_1)$.
\noindent 
(2) Even if $j_1 = j_2$, there is a difference
between assigning two different spin states $m_1$ and $m_2$ to the two different
edges.

In calculating the entropy of equation (\ref{final_entropy}), the edges were 
considered to be {\it completely distinguishable}, i.e. both claims (1) and (2) 
are imposed. Since the positions of different edges are determined by the 
{\it spin network}, which is formed by the way in which the edges are connected 
to each other and to the outside world, this viewpoint appears to be the one best describing the 
physical model, and as such, the one we should adopt (see also \cite{strom}).

In papers \cite{entropy,CM} a different view is adopted: The edges are {\it partially
distinguishable}. Indeed, the multiplicity formula 
$\Omega (\{N_n\})=\prod_n g(n)^{N_n}$ applies if claim (1) above is {\it not}
adopted while (2) still applies. 
Evaluating (\ref{partitionfcn}) with this counting
of states we obtain
\begin{equation} \label{partitionfcn3}
  Z = \sum_{\{N_n\}} \Omega(\{N_n\}) e^{-\beta A} = \prod_n \frac{1}
{1-g(n) e^{-\beta A(n)}}
\end{equation}
If each edge does not have an exponentially increasing density of states (in which
case it would be itself a macroscopic black hole) the only poles of the above
expression are at $e^{\beta A(n)} = g(n)$. Calling ${\bar \beta}_o$ the lowest
of these values of $\beta$, occuring for some $n_o$, we deduce that for 
macroscopic areas the model will exhibit Bose condensation at the spin
$j_o = (n_o -1)/2$ and the entropy will be $S= {\bar \beta}_o A$.
For our area function (\ref{Aj}) and the degeneracy function $g(n)=n$ this gives an 
entropy $S=\frac{kA\gamma_0}{4 l_p^2}$ with $\gamma_0=\frac{\ln 3}{3\pi}$.
Bose condensation occurs at spin $j=1$.
For Ashtekar's choice of area function the result is instead 
$\gamma_0=\frac{\ln 2}{\pi\sqrt{3}}$, and Bose condensation occurs at 
$j=\frac{1}{2}$.

Finally, if the edges are {\it identical} we adopt neither claim (1) nor (2)
and have the equivalent of a Bose-Einstein gas. Then, it turns out that the entropy
is related to the area by $S \propto  A^{t}$, where the exponent satisfies $t<1$ if 
each edge does not have an exponential density of states. For the choices of area 
spectrum (\ref{Aj}) and degeneracy function $g(n)=n$, the exponent acquires
the value $t=\frac{2}{3}$. No Bose condensation occurs.

In conclusion, the area law $S = \beta_o A$ and the appearence of a Hagedorn
temperature are the main results. These are extremely generic, requiring only
{\it some} distinguishability of the elementary components. 
Since the Immirzi parameter is not fixed by the quantum theory \cite{Immirzi}, 
the different values of $\beta_o$ are of somewhat secondary importance. 
There is,
however, a crucial physical difference between our result and the result of
\cite{entropy,CM}. While the latter suggest that Bose condensation occurs
and area fluctuations are strongly suppressed, 
in our fully distinguishable case no Bose condensation occurs
and the area within any solid angle of the black hole will exhibit 
fluctuations of order $\sqrt A$.

{\bf Acknowledgements}. We thank J. Baez, V. Mukhanov and T. Strobl
for useful discussions.

\def\baselinestretch{1.0}

\widetext


\begin{references}
\small

\bibitem{Gambinibook} R. Gambini, J. Pullin, 
{\it Loops, Knots, Gauge Theories and Quantum Gravity}
Cambridge University Press, Cambridge, 1996.
 
\bibitem{Ashtekarbook} A. Ashtekar, 
{\it Lectures on  non-perturbative canonical gravity}
Advanced Series in Astrophysics and 
Cosmology, Vol.6, World Scientific, Singapore, 1991. 


\bibitem{D0} T. Banks {\em et al.}  JHEP {\bf 9801} (1998) 008;
H. Liu, A. A. Tseytlin JHEP {\bf 9801} (1998) 010.


\bibitem{Immirzi} C. Rovelli and T. Thiemann, Phys.Rev. 
{\bf D57} (1998) 1009-1014.

\bibitem{RS_spectrum} C. Rovelli and  L. Smolin, Nucl.Phys. {\bf B442} (1995) 593-622;
S. Frittelli, L. Lehner and C. Rovelli, 
Class.Quant.Grav. {\bf 13} (1996) 2921-2932.

\bibitem{spectroscopy} J. D. Bekenstein, V. F. Mukhanov, Phys.Lett. {\bf B360} (1995) 
7-12.

\bibitem{entropy} A. Ashtekar {\em et al.} Phys.Rev.Lett. {\bf 80} 
(1998) 904-907.

\bibitem{CM} S. Chaudhuri and  D. Minic, Phys. Lett. 
{\bf B433} (1998) 301-306.


\bibitem{Baez} J.C.Baez, {\it An Introduction to Spin Foam Models 
of BF Theory and Quantum Gravity} preprint gr-gc/9905087.


\bibitem{Duf} M. Duflo, Ann. Sci. \'Ecole Norm. Sup. {\bf 10} (1977)
265-288.

\bibitem{NR} H.B. Nielsen and D. Rohrlich, Nucl. Phys. {\bf B299} (1988) 471. 

\bibitem{AFS} A. Alekseev, L. Faddeev and S. Shatashvili, J. Geom. Phys.
{\bf 5}, no. 3 (1989) 391.

\bibitem{Witten} Edward Witten, J.Geom.Physics {\bf 9} (1992) 303-368.

\bibitem{MW} E. Meinrenken and C. Woodward, Progr. Math. {\bf 172} (1999)
271-295.

\bibitem{Immirzi} C. Rovelli and T. Thiemann, Phys.Rev. 
{\bf D57} (1998) 1009-1014.

\bibitem{strom} A. Strominger,
Phys. Rev. Lett. {\bf 71} (1993) 3397-3400.









\end{references}
\end{document}